\documentclass[
aps,
reprint,
pra,
amsmath,amssymb,
]{revtex4-1}

\usepackage[utf8]{inputenc}
\usepackage[english]{babel}

 \usepackage{graphics}
 \usepackage{graphicx}
 \usepackage{longtable}
 \usepackage{multirow}
 \usepackage{url}
 \usepackage{bm}
 \usepackage{color}
 \usepackage[dvipsnames]{xcolor}
 \usepackage[euler]{textgreek}
 \usepackage[normalem]{ulem}

 \usepackage{paralist}

\begin{document}
\title{Deeply bound (24$D_J$ + 5$S_{1/2}$) $^{87}$Rb and $^{85}$Rb molecules for eight spin couplings}
\author{Jamie L. MacLennan}
\author{Yun-Jhih Chen}
 \altaffiliation[Present address:]{ National Institute of Standards and Technology, Boulder, Colorado 80305, USA}
\author{Georg Raithel}
\affiliation{Department of Physics, University of Michigan, Ann Arbor, MI 48109}
\date{\today}

\begin{abstract}
We observe long-range $^{85}$Rb and $^{87}$Rb (24$D_{J}+$5$S_{1/2}$) Rydberg molecules for eight different spin couplings, with binding energies up to 440~MHz and subpercent relative uncertainty. Isotopic effects of the molecular binding energies arise from the different masses and nuclear spins. Because the vibrational states involve different spin configurations and cover a wide range of internuclear separations, the states have different dependencies on the \mbox{$s$-wave} and \mbox{$p$-wave} scattering phase shifts for singlet and triplet scattering. Fitting the spectroscopic data, we comprehensively determine all four scattering length functions over the relevant energy range as well as the zero-energy scattering lengths of the two \mbox{$s$-wave} channels. Our unusually high temperature and low density (180~$\mu$K, 1~$\times$~10$^{11}$~cm$^{-3}$) suggest that the molecule excitation occurs through photoassisted collisions.
\end{abstract}
\maketitle

\section{Introduction}
The scattering of a Rydberg electron and a neutral ground-state atom is a unique mechanism of forming a molecular bond \cite{Greene2000prlcreation}, which is fundamentally different from covalent, ionic, or van der Waals bonds. Experimentally accessible characteristics of these ``Rydberg-ground'' molecules, such as vibrational energy levels and dipole moments, depend on the electron-atom scattering phase shifts in the sub-50-meV range. Measurements of their molecular binding energies can validate calculations of the scattering phase shifts and the structure of negative-ion resonances \cite{Fabrikant1986jpbinteraction,Bahrim2000pralowlying,Bahrim2001jpb3Seand1Se,Fabrikant2001pranegativeion}. Studying low-energy-electron scattering using electron and molecular beams is difficult due to inherent energy spreads and space-charge electric fields. Rydberg molecules present an attractive, experimentally accessible alternative~\cite{DeSalvo2015praultralongrange, Merkt2015prlexpchar, Bendkowsky2009natobservationultralong, Bendkowsky2010prlrydbergtrimers, Anderson2014prlphotoassociation, Pfau2016praobservation, Krupp2014prlalignment, Tallant2012prlobservationblue, Bellos2013prlexcitation, Niederpruem2016prlspinflips}, in which electric fields can be eliminated using Rydberg Stark spectroscopy~\cite{Gallagher1994}. Thus these molecules emerge as a testbed for low-energy electron-atom scattering \cite{Stebbings1983bookrydbergstates, Klar1994zpdcomparison, Dunning1995jpbemcollisions, Fabrikant1996aipemscattering, Khuskivadze2002praadiabatic, Hotop2003advresonance}. Low-energy electron scattering is also of broad interest. For instance, it can cause DNA strand breaks through the formation of negative-ion resonances \cite{Alizadeh2015apcbiomolecular, Bald2006acieselectiveexcision, Simons2006acshowdo, Martin2004prldnastrand, Caron2003prllowenergyel}.

\section{Rydberg-Ground Molecules with Hyperfine Structure}
The Rydberg-ground molecular interaction may be described by a Fermi pseudopotential \cite{Fermi1934incsopra, Omont1977jponthetheory} in which the ground-state atom is modeled as a point perturber. The perturbation strength is determined by energy-dependent scattering lengths $a_l(k)$, which are related to the scattering phase shifts $\eta_l(k)$ by $a_l(k)^{2l+1} = - \tan \eta_l(k) / k^{2l+1}$, where $k$ is the electron momentum and $l$ is the scattering partial-wave order ($s$, $p$, ...). In the reference frame of the Rydberg ionic core, the scattering interaction is \cite{Omont1977jponthetheory}:
\begin{equation}
\begin{split}
\label{eqn:V}
\hat{V}(\mathbf{r}; R)& = 2\pi a_s(k)\delta^3(\mathbf{r}-R\hat{\mathbf{z}}) \\
& \quad + 6\pi [a_p(k)]^3\delta^3(\mathbf{r}-R\hat{\mathbf{z}}) \overleftarrow{\nabla} \cdot \overrightarrow{\nabla}\
\end{split}
\end{equation}
where $\mathbf{r}$ and $R\hat{\mathbf{z}}$ are the positions of the Rydberg electron and perturber atom. Previous measurements of vibrational energies of low-angular-momentum diatomic Rydberg-ground molecules have spanned principal quantum numbers $n$ = 26-45, angular momentum $S$, $P$, and $D$ states, and atomic species rubidium, cesium, and strontium \cite{DeSalvo2015praultralongrange, Merkt2015prlexpchar, Bendkowsky2009natobservationultralong, Bendkowsky2010prlrydbergtrimers, Anderson2014prlphotoassociation, Pfau2016praobservation, Krupp2014prlalignment}. For Sr, the \mbox{$s$-wave} and \mbox{$p$-wave} zero-energy scattering lengths, $a_s(0)$ and $a_p(0)$, were extracted from \mbox{$S$-state} data \cite{DeSalvo2015praultralongrange}. In Rb and Cs electron-scattering, there are two relevant electrons. For Cs, the corresponding singlet and triplet s-wave scattering lengths, $a_s^S(0)$ and $a_s^T(0)$, were extracted from mixed singlet-triplet resonances in \mbox{$P$-states} \cite{Merkt2015prlexpchar} using a model developed in \cite{Anderson2014praangmom}. In Rb, $a_s^T(0)$ was extracted from $S$- and \mbox{$D$-state} molecular resonances \cite{Bendkowsky2009natobservationultralong, Bendkowsky2010prlrydbergtrimers, Anderson2014prlphotoassociation} and $a_p^T(0)$ from $S$-state resonances \cite{Bendkowsky2010prlrydbergtrimers}. Mixed singlet-triplet resonances in Rb \mbox{$S$-states} \cite{Pfau2016praobservation} allowed an extraction of $a_s^S(0)$ after determining $a_s^T(0)$ from previous data \cite{Bendkowsky2009natobservationultralong, Krupp2014prlalignment}. To our knowledge, $a_p^S(k)$ at any $k$ has not been measured.

Here, we measure the binding energies of 24$D_{J}-$5$S_{1/2}$ $^{85}$Rb and $^{87}$Rb molecular states for eight combinations of spin couplings, with fractional uncertainties as low as 0.2\% for the deepest states. Unique sets of resonances for each combination reveal the dependence on the isotopic mass and, notably, the nuclear spin $I_2$ of the ground state atom. We fit the binding energies to 3.8~MHz rms deviation with a semi-empirical model and extract scattering length functions for all four scattering channels, including the singlet $p$-wave one. We discuss why our relatively hot (temperature $180~\mu$K) and dilute (density $\gtrsim 10^{11}$~cm$^{-3}$) atom sample yields a surprisingly strong molecular signal.

\begin{figure}[t]
\centering
\vspace*{-0.18in}
\includegraphics[width = \columnwidth]{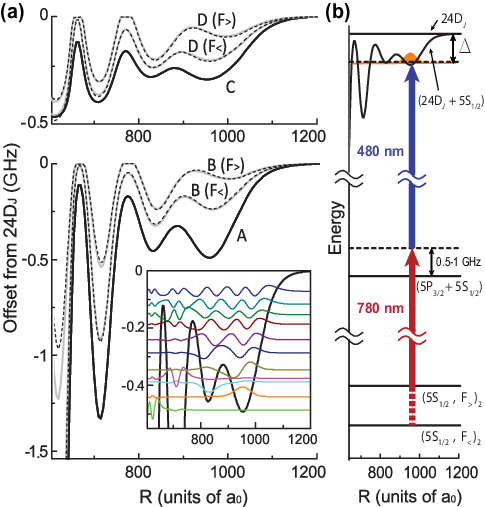}
\setlength{\abovecaptionskip}{-1ex}
\setlength{\belowcaptionskip}{-3ex}
\caption{(Color online.) (a) Potential curves for Rb (24$D_{J}$ + 5$S_{1/2}$) molecules for J=5/2 (top) and J=3/2 (bottom). The ``deep'' potentials (solid black) are virtually the same for both hyperfine ground-states (F$_>$ and F$_<$) and isotopes ($^{87}$Rb and $^{85}$Rb). The ``shallow'' potentials (solid gray for $^{87}$Rb, and dashed black for $^{85}$Rb) depend significantly on hyperfine ground-state and slightly on isotope. Inset shows wavefunctions of vibrational resonances in potential~A (vertical offset shows resonance energy). (b) Excitation level diagram.}
\label{fig:pecs_ex_joined}
\end{figure}

The full Hamiltonian for the system is~\cite{Anderson2014praangmom}:
\begin{eqnarray}
\label{eqn_hamiltonian}
\hat{H}(\mathbf{r}, R)=&&\hat{H}_0 + \sum_{i=S,T} \hat{V}_i(\mathbf{r}, R) \mathcal{\hat{P}}_i + A_{\text{HFS }} \hat{\mathbf{S}}_2 \cdot \hat{\mathbf{I}}_2 \
\end{eqnarray}
where $\hat{H}_0$ is the Hamiltonian of the unperturbed Rydberg electron (including its fine structure). The second term sums over both spin-dependent singlet ($i$=S) and triplet ($i$=T) scattering channels, using the projection operators $\mathcal{\hat{P}}_T = \hat{\mathbf{S}}_1 \cdot \hat{\mathbf{S}}_2 + 3/4$, $\mathcal{\hat{P}}_S = 1- \mathcal{\hat{P}}_T$ ($\hat{\mathbf{S}}_1$ and $\hat{\mathbf{S}}_2$ are the electronic spins of the Rydberg and ground-state atom, respectively). The last term represents the hyperfine coupling of $\hat{\mathbf{S}}_2$ to the ground-state-atom nuclear spin $\hat{\mathbf{I}}_2$, with hyperfine parameter $A_{\text{HFS}}$. In Rb, $A_{\text{HFS}}$ is comparable to the scattering interactions (on the order of GHz), and $\hat{\mathbf{I}}_2$ becomes coupled in second order to $\hat{\mathbf{S}}_1$ through $\mathcal{\hat{P}}_T$ and $\mathcal{\hat{P}}_S$. The singlet potentials disappear and are replaced with mixed singlet-triplet potentials \cite{Anderson2014prlphotoassociation, Anderson2014praangmom}. These, in addition to the (nearly-pure) triplet potentials,
sustain molecular bound states, as has been observed in Cs \cite{Merkt2015prlexpchar} and Rb \cite{Pfau2016praobservation, Niederpruem2016prlspinflips}.

\begin{figure}
\centering
\includegraphics{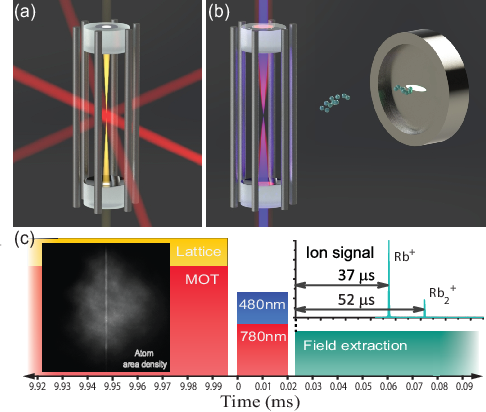}
\setlength{\belowcaptionskip}{-3ex}
\caption{(Color online.) Experimental sequence: (a) Atoms are first trapped in a MOT (red beams) and loaded into the vertical 1-D lattice trap (yellow). (b) The traps are switched off, and overlapping 780-nm and 480-nm beams excite a Rydberg atomic or molecular state. After excitation, voltages applied to six metal rods steer spontaneously generated Rb$^+$ and Rb$_2^+$ ions to the MCP detector, where they arrive in time-resolved clusters. (c) Timing sequence. Data rate is 100 Hz. The insets show a qualitatively-representative atom area density of the lattice-trapped atoms and surrounding MOT (left) and an ion time-of-flight signal (upper right).}
\label{fig:cavity}
\end{figure}

We obtain the potential energy curves (PECs) by solving the Hamiltonian on a grid of intermolecular distances $R$, \cite{Greene2000prlcreation, Bendkowsky2009natobservationultralong}, as shown in Fig.~\ref{fig:pecs_ex_joined}(a). Following the Born-Oppenheimer approximation, the PECs describe the vibrational motion. The hyperfine-mixed singlet-triplet potentials (``shallow'' potentials) have shallower wells and vary significantly depending on whether the ground-state-atom is in its upper or lower hyperfine state, $F_2 = F_>$ or $F_<$. The shallow potentials for $F_<$ are deeper than those for $F_>$. The triplet potentials (``deep'' potentials) are virtually unaffected by hyperfine mixing, and therefore independent of $I_2$ and $F_2$.

The narrow molecular resonances in each PEC are found by solving the Schr\"odinger equation for the vibrational motion \cite{Anderson2014praangmom}. The result is a spectrum of vibrational states, the majority of which are mostly contained in the outermost potential wells [inset of Fig.~\ref{fig:pecs_ex_joined}(a)].

\begin{figure*}
\centering
\includegraphics[width = \textwidth]{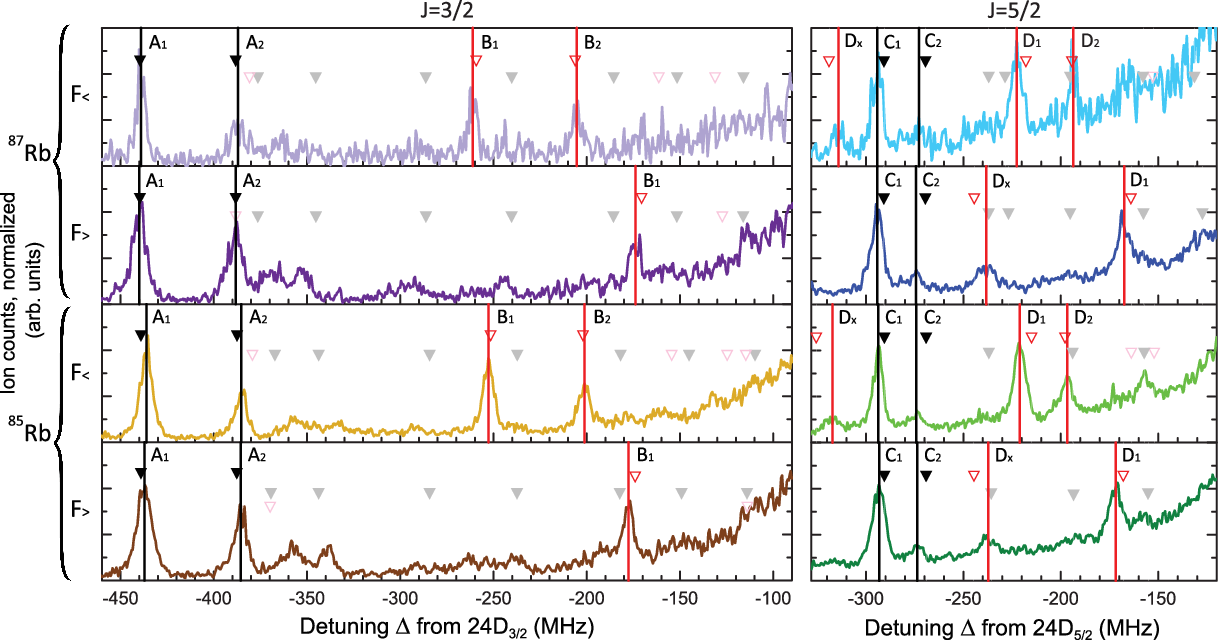}
\setlength{\abovecaptionskip}{-1ex}
\setlength{\belowcaptionskip}{-1ex}
\caption{(Color online.) Detected ions vs. detuning relative to the 24$D_{J}$ atomic state, for the eight $(I_2, F_2, J)$-combinations. The spectra are normalized by the height of the A$_1$ or C$_1$ resonance. A selection of resonances is marked with vertical lines and labeled according to their corresponding potential in
6
 Fig.~\ref{fig:pecs_ex_joined}(a). Filled (open) triangles denote resonances in the deep (shallow) potentials predicted with our model. Faded triangles are additional predicted resonances not used in the fitting procedure.}
\label{fig:data}
\end{figure*}

\begin{table*}
\begin{ruledtabular}
\small
\begin{tabular}{c c c c c c c c c c}
Pair~potentials & A$_1$ & A$_2$ & B$_1$ &B$_2$ & C$_1$ & C$_2$ & D$_x$ & D$_1$ & D$_2$\\
 \hline
 $^{87}$Rb (24$D_J$ + 5$S_{1/2}$ F$_{<}$) &-439.1(10)&-387.0(10)&-261.2(8)&-205.4(8)&-294.1(8)&-272.7(8)&-314.0(8)&-222.6(7)&-193.6(7)\\
 $^{87}$Rb (24$D_J$ + 5$S_{1/2}$ F$_{>}$) &-439.9(10)&-388.2(10)&-173.9(7)& &-294.1(8)&-274.2(7)&-238.3(7)&-167.5(7)& \\
 $^{85}$Rb (24$D_J$ + 5$S_{1/2}$ F$_{<}$) &-436.1(9)    &-385.3(9)&-252.7(8)&-201.2(7)&-293.7(8)&-274.1(8)&-317.2(8)&-221.0(7)&-196.7(7)\\
 $^{85}$Rb (24$D_J$ + 5$S_{1/2}$ F$_{>}$) &-437.2(9)    &-385.5(9)&-177.7(7)& &-293.2(8)&-273.7(8)&-237.2(7)&-171.8(7)& \\
\end{tabular}
\setlength{\abovecaptionskip}{-0.1ex}
\setlength{\belowcaptionskip}{-2ex}
\caption{Molecular binding energies in MHz, relative to the atomic lines, corresponding to the labeled peaks in Fig.~\ref{fig:data}.}
\label{fig:binding_energy_table}
\end{ruledtabular}
\end{table*}

\section{Experimental Setup}
In the experiment we photoassociate Rydberg molecules from cold Rb atoms out of a one-dimensional (1D) lattice dipole trap, which is loaded from a magneto-optical trap (MOT). An atom pair undergoes two-photon excitation to a (24$D_{J} + $5$S_{1/2}$) pair state [Fig.~\ref{fig:pecs_ex_joined}(b)]. The fixed lower-transition (780-nm) frequency is 0.5-1~GHz blue-detuned from the $5P_{3/2}$ intermediate state to mitigate scattering-induced heating, while the upper-transition (480-nm) frequency is scanned from the Rydberg atomic line to several hundred MHz below. Rydberg molecules are formed when the detuning from the atomic line matches a molecular binding energy.

We prepare molecules for eight cases of ($I_2$, $F_2$, $J$): (i) To observe the isotope ($I_2$) dependence, we adjust our MOT lasers to trap either isotope. (ii) To observe the hyperfine ($F_2$) dependence, we prepare the atoms in either $F_>$ or $F_<$ by turning off the repumper either at the same time as the cooling laser or 150 $\mu$s earlier. We adjust the 780-nm laser frequency according to our choice for (i) and (ii). Finally, (iii) to observe 24$D$ $J = 3/2$ or $5/2$ Rydberg-ground molecules, the 480-nm laser frequency is changed by the Rydberg fine-structure splitting (913 MHz).

Figure~\ref{fig:cavity} shows the experimental geometry and timing. The lattice trap (1064 nm) is formed by a fundamental Gaussian mode of a near-concentric, in-vacuum cavity \cite{Chen2014praatomtrapping} with a trap depth of $\sim$40~MHz for Rb 5S$_{1/2}$. The lattice trap loads $\sim$2$ \times 10^4$ atoms from an overlapping MOT (Fig.~\ref{fig:cavity}(a)) to generate an atom cloud of about 18~$\mu$m diameter, 700~$\mu$m length, transverse temperature $\sim$180~$\mu$K, and central volume density $\sim$1.6$ \times 10^{11}$ cm$^{-3}$. Before photoassociation, we turn off the MOT and the lattice trap to avoid light shifts. Several $\mu$s afterward, a 20-$\mu$s pulse of 780-nm and 480-nm light excites atoms to Rydberg atomic and molecular states. The 780-nm and 480-nm beams have respective waists of 20~$\mu$m and $\lesssim$ 100~$\mu$m and overlap with the 1064-nm trap, creating an oblong excitation volume in the densest region of the cloud.

The Rydberg-ground molecules yield either Rb$_2^+$ via Hornbeck-Molnar autoionization or ion-pair formation, or Rb$^+$ via black-body photoionization or ion-pair formation \cite{Cheret1982jpbpenning, Barbier1987expstudypenning, Bendkowsky2010phdthesis, Ott2015prlgiant}. The ions detected by the microchannel plate [MCP; see Fig.~2(b)] are our signal. Rydberg-Rydberg molecules are also produced \cite{Zhao2018praCsnDJ, Boisseau2002prlultralong, Sassmannshausen2016prlobservation}, but only a small fraction ionize spontaneously. Since the ion extraction electric field is too weak to field-ionize the Rydberg products, we preferentially detect Rydberg-ground molecules.

\section{Experimental Results}
The resonances in the spectra shown in Fig.~\ref{fig:data} are different
for each ($I_2$, $F_2$, $J$) case. The upward trend of the background signal at small detunings is attributed to Rydberg-Rydberg molecules~\cite{Zhao2018praCsnDJ}. We assign the most prominent peaks to the deep or shallow potentials of Rydberg-ground molecules, A/C or B/D in Fig.~\ref{fig:pecs_ex_joined}, respectively, by comparison with the resonances predicted by our model (see below). The binding energies, determined by Gaussian fits to the peaks, are marked with vertical lines and listed in Table \ref{fig:binding_energy_table}. The largest uncertainty arises from the 480-nm laser frequency calibration (typically 0.6-0.9~MHz), followed by statistical uncertainties caused by fluctuations in signal strength and the peak-fitting. The resultant relative uncertainties of the listed binding energies range between 0.2\%, for the lowest states found, and 0.4\%. To our knowledge, these values are lower than previously reported ones.

The identified peaks, with the exception of peaks D$_x$, arise from the first or second vibrational resonances in the outer region of the PECs ($R \approx$ 800--1000$a_0$, see Fig.~\ref{fig:pecs_ex_joined}). We observed no other prominent resonances up to 150~MHz below the deepest peaks in Fig.~\ref{fig:data}. The D$_x$ peaks correspond to resonances in the inner well at $R \approx 710a_0$; they have weaker signals due to the reduced likelihood of finding atoms at smaller internuclear separations. Most unidentified peaks in Fig.~\ref{fig:data} are higher resonances in the deep potentials. Their low signal strength may be attributed to the rapid oscillations in their vibrational wavefunctions [see the inset of Fig.~\ref{fig:pecs_ex_joined}(a)], leading to small Franck-Condon factors, and their short tunneling-induced lifetimes. The spin-mixing also plays a role in signal strength; a method for computing line strengths of vibrational spectra of Rydberg molecules including the hyperfine- and spin-dependence is presented in Ref. \cite{Markson2016cpctheory}.

The starkest difference among the spectra in Fig.~\ref{fig:data} is between $J$=5/2 (right) and $J$=3/2 (left); the deepest resonances differ by up to 150~MHz. The depths of the molecular potential wells and the fine structure scale as $n^{-6}$ \cite{Krupp2014prlalignment} and $n^{-3}$, respectively. At low $n$, the molecular binding interaction strength exceeds the fine-structure splitting. When this happens, the molecules are classified as Hund's case (a) \cite{hund1927,brown2003}. In this limit, the molecular potentials that asymptotically connect to $J$=5/2 approach and repel from the adiabatic potentials that connect to the $J$=3/2 atomic level. As a result, in the Hund's case (a) regime, the $J$=3/2 adiabatic potentials become deeper than the fine-structure coupling, with their depths scaling as $n^{-6}$, whereas the $J$=5/2 potentials are limited in depth by the fine structure splitting; hence their depths scales as $n^{-3}$. Molecules in Rb 24$D_{J}$ are far into the Hund's case (a) regime. The relevance of Hund's cases to Rydberg-ground molecules has been discussed in detail in Ref. \cite{Anderson2014praangmom}.

The largest difference among the rows in Fig.~\ref{fig:data} and Table \ref{fig:binding_energy_table} is between the states in the shallow potentials (i.e., the $B$ and $D$ peaks) for $F_<$ and $F_>$, which differ by up to 70~MHz. The strong dependence on $F_2$ is expected from the PECs in Fig.~\ref{fig:pecs_ex_joined}. The $B$ and $D$ peaks also exhibit isotopic differences up to $\sim$10~MHz, which originate from the different hyperfine-coupling strengths $A_{\text{HFS}}$, nuclear spins $I_2$, and masses.

The $A$ peaks are similar for $F_>$ and $F_<$ but vary slightly between the two isotopes (see Fig.~\ref{fig:data} and Table~\ref{fig:binding_energy_table}). As the $A$-PECs are virtually identical, the variation is likely due to the isotopic mass difference. The heavier isotope has deeper binding energies because of its smaller vibrational frequencies within the same potential. The two unlabeled resonances immediately to the right of $A_{2}$ (at -370 to -330~MHz) show a pronounced difference between isotopes, suggesting that they correspond to states of the inner PEC wells at $R$ $\approx$ 710$a_0$ in Fig.~\ref{fig:pecs_ex_joined}, where a mass difference has a greater quantitative effect because of the larger spacings between vibrational states.

\section{Model and Fitting Procedure} 
To model the observed molecular resonances based on Eqs.~1 and~2, four scattering-length functions $a_l^i(k)$ are required. In our model, we use the short-range potential provided in Ref.~\cite{Khuskivadze2002praadiabatic}, integrate the radial Schr\"odinger equation, and evaluate the scattering wavefunctions at a distance $d =$ 150$a_0$ from the perturber, corresponding to the typical width of the outermost lobe of the Rydberg-electron wavefunction for 24$D$. The scattering lengths then follow from textbook equations \cite{Sakurai2010}. The values of the resulting scattering length functions $\tilde{a}_l^i(k)$  at very low $k$ are artificial because they depend on the evaluation distance (here, $d$ = 150$a_0$), whereas the true scattering length functions $a_l^i(k)$, obtained in the limit $d \rightarrow \infty$, are independent of $d$. Our approach of using $d$ = 150$a_0$ avoids the problem that for $k$ $\rightarrow 0$ the \mbox{$p$-wave} scattering lengths diverge \cite{OMalley1961jmpmodification}, which would cause an unphysical divergence in the adiabatic potentials at the classical turning point of the Rydberg electron when using the Fermi method. Due to the localization of the Rydberg electron within the lobes of its wavefunction, the probability of finding it at very low $k$ is negligible, allowing us to use $\tilde{a}_l^i(k)$ to calculate the potentials.

\begin{table}[ht]
\begin{ruledtabular}
\begin{tabular}{c c c c}
 & $a_s^T(0)$ & $a_s^S(0)$ & Ref.\\
 \hline
Theory~~~~~~ & -16.1~~~~  & 0.627 & \cite{Bahrim2001jpb3Seand1Se}\\
& -16.9~~~~ & 0.63 & \cite{Eiles2018praformation}\\
\hline
Experiment & -15.7(1) & neglected & \cite{Krupp2014prlalignment}\\
&~~-15.7(1)* & -0.2(5) & \cite{Pfau2016praobservation}\\
& -14.0(5) & neglected & \cite{Anderson2014prlphotoassociation}\\
& -14.7(3) & ~0.0(3) & This work\\
\end{tabular}
\setlength{\abovecaptionskip}{-0.1ex}
\setlength{\belowcaptionskip}{-1ex}
\caption{Zero-energy scattering lengths in $a_0$.\\
*$a_s^T(0)$ was fixed while $a_s^S(0)$ was fitted.}
\label{fig:scat_length_table_all}
\end{ruledtabular}
\end{table}

In our fitting procedure, we allow for adjustable phase shifts of the scattering wavefunctions at 0.01$a_0$, near the center of the perturber atom, which account for short-range corrections of the Rb$^-$ scattering potentials and are used to fine-tune the functions $\tilde{a}_l^i(k)$. Every set of four $\tilde{a}_l^i(k)$ yields eight PECs through solving Eq.~\ref{eqn_hamiltonian}. From the PECs we obtain the vibrational resonances and determine their rms deviation from the 32 measured values in Table~\ref{fig:binding_energy_table}. The four adjustable phases are varied and the procedure is repeated until the rms deviation is minimized (3.8~MHz). The corresponding calculated resonances are shown as triangles in Fig.~\ref{fig:data}.

\begin{figure*}[ht]
\centering
\includegraphics[width = \textwidth]{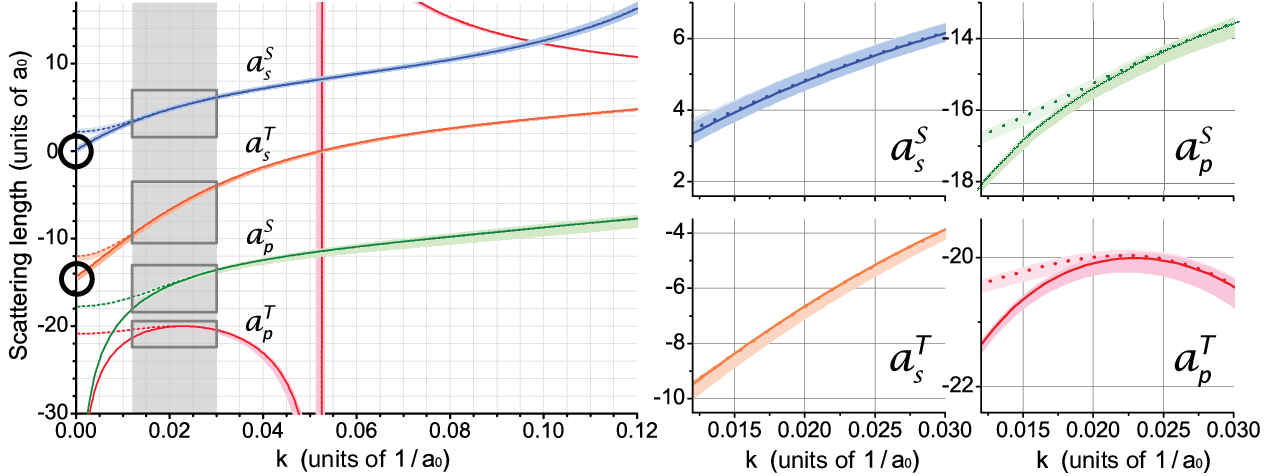}
\setlength{\abovecaptionskip}{-2ex}
\setlength{\belowcaptionskip}{-1ex}
\caption{(Color online.) Scattering length functions for the $a_l^i(k)$ (solid lines; $d = 2 \times 10^4a_0$) and $\tilde{a}_l^i(k)$ (dashed lines; $d = 150a_0$) that correspond to the predicted resonances in Fig.~\ref{fig:data}. Shaded backdrops behind the curves show the uncertainties. Vertical gray strip corresponds to the experimentally relevant energy range; the four inscribed rectangles correspond to the zoom-ins shown in the four panels on the right. Black circles indicate the two zero-energy values included in Table~\ref{fig:scat_length_table_all} for this work.}
\label{fig:scat_length_combo}
\end{figure*}

To estimate the zero-energy values of the true scattering length functions, we also calculate the functions ${a}_s^i(k)$ using an evaluation distance $d$ = $2 \times 10^4a_0$. Figure~\ref{fig:scat_length_combo} shows the four extracted scattering length functions $\tilde{a}_l^i(k)$ and their corresponding ${a}_l^i(k)$. As expected, $\tilde{a}_l^i(k)$ and $a_l^i(k)$ match at $k \gtrsim 0.015$ ($E = \hbar^2 k^2 / 2m \gtrsim 3$ meV). We anticipate the predicted scattering lengths to be the most useful in the range $k = 0.012-0.030$ (shaded vertical strip in Fig.~\ref{fig:scat_length_combo}), which corresponds to $E = 2-12$ meV and $R = 700-960 a_0$, because this is the region probed by the measured molecular bound states.

We quantify the uncertainty in $\tilde{a}_l^i(k)$ and $a_l^i(k)$ by varying several parameters in our procedure. First, we perform the fitting procedure for three Rydberg-state basis sets $21.1 - j \leq n\mbox{*} \leq 24.1 + j$, with effective principal quantum number $n\mbox{*}$, for $j =$ 0, 1 and 2. In Fig.~\ref{fig:data}, the $j = 2$ basis size is used. Secondly, we include or omit resonances in the inner potential well at $R = 710a_0$ (see Fig.~\ref{fig:pecs_ex_joined}). Thirdly, we increase or decrease the measured resonance values by the experimental frequency uncertainty ($\sim0.2\%$). We use the combination of the three sources as the estimated uncertainty (in Fig.~\ref{fig:scat_length_combo} and Table~\ref{fig:scat_length_table_all}).

We are able to simultaneously fit the four scattering lengths because we analyze a large set of binding energies on PECs for a variety of spin cases, which have different sensitivities to the singlet and triplet \mbox{$s$-wave} and \mbox{$p$-wave} scattering-length functions. For instance, the inner-well resonances at 710$a_0$ (e.g., $D_x$) and 820$a_0$ (e.g., $A_2$, $B_1$, $C_2$, $D_1$) depend strongly on $\tilde{a}_p^T(k)$ (whose shape resonance is responsible for the steep drop-off in Fig.~\ref{fig:pecs_ex_joined}(a) at 600$a_0$) and weakly on $\tilde{a}_s^T(k)$, while outer-well resonances show the opposite trend. Meanwhile, substantial dependencies on $\tilde{a}_s^S(k)$ and $\tilde{a}_p^S(k)$ are only found in the B and D resonances. Some small codependencies of the scattering lengths remain. For instance, fixing $\tilde{a}_s^S(k)$ would significantly decrease the uncertainty in $\tilde{a}_p^S(k)$.

We are quoting an experimental result for $a_p^S(k)$ [and for $a_l^i(k)$ for $E = 2-12$ meV] derived from a fitting procedure applied in a Rydberg-ground molecule experiment. In Table~\ref{fig:scat_length_table_all}, we show our median values (within the uncertainty bounds) of ${a}_s^T(0)$ and ${a}_s^S(0)$ for comparison with other published zero-energy values.

\section{Discussion}
We note several deficiencies of the method we have used. First, the choice of basis size used to calculate the adiabatic potentials affects the depth of the potentials. We explored the convergence behavior of the adiabatic potentials as a function of basis size, ranging from $21.1 \leq n\mbox{*} \leq 24.1$ to $17.1 \leq n\mbox{*} \leq 28.1$ (i.e. we varied the range in $n\mbox{*}$ from about 3 to 11). Over this substantial variation in $n\mbox{*}$ range, we found that the outer potential wells increased in depth by 13\% over the entire range, and that they do not seem to converge with growing basis size (but the incremental changes decrease). This is problematic and raises the question of which basis choice leads to the most accurate potentials. The issue of non-convergence has also been noted elsewhere and discussed in comparison to alternative techniques for calculating the adiabatic potentials \cite{Eiles2017prahamiltonian, Hamilton2002jpbshape, Chibisov2002jpbenergies, Fey2015njpcomparitive}, and the topic has been described as controversial. A second deficiency of our method is that the Fermi model may have fundamental inaccuracies at low $n$, where the size of the perturber atom relative to the Rydberg wavefunction increases. This could possibly be addressed by using a Green's function calculation \cite{Khuskivadze2002praadiabatic}.

The minor discrepancies between our quoted zero-energy scattering lengths and previous results have several possible causes. Methods for calculating the $k$-dependence of $a_l^i(k)$ vary. The inaccuracy of the Fermi model at low $n$ may contribute. We also note that the previously-quoted experimental scattering lengths did not account for \mbox{$p$-wave} scattering, which may have caused the extracted \mbox{$s$-wave} values to be overly negative. Finally, we note that in Ref. \cite{Bendkowsky2010prlrydbergtrimers} two values for $a_s^T(0)$ are presented, -16.05$a_0$ and -19.48$a_0$, along with a zero-energy \mbox{$p$-wave} scattering length $a_p^T(0)$ of -21.15$a_0$.

The reported results are obtained with atoms prepared at a temperature of $\sim$180~$\mu$K, densities of only $\gtrsim10^{11}$~cm$^{-3}$, and a quantum state as low as $n$ = 24. The strongest molecular signal is about $1\%$ of the signal on the atomic line (not shown in Fig.~\ref{fig:data}). This is surprising because under the given conditions the instantaneous probability of finding a ground-state atom within a Rydberg atom is only about $0.01\%$. This discrepancy may be resolved by interpreting the molecule excitations as photo-assisted collisions. Estimates show that the excitation pulse duration, Rydberg-excitation Rabi frequency, thermal velocities and atom density are such that during the excitation pulses the fraction of atom pairs that undergo collisons at distances of the typical vibrational bond length ($\sim$800$a_0$) is sufficient to explain the molecular-signal strength. In a photo-assisted collision, in contrast to the concept of a frozen Rydberg gas, the molecule excitation can be considered a non-adiabatic transition of atomic wave-packets between intersecting dressed-atom ground-ground and ground-Rydberg PEC's that are coupled by the Rydberg Rabi frequency. Further analysis of this scenario is ongoing.

\section{Conclusion}
In summary, we have measured 32 binding energies of (24$D_{J}+$5$S_{1/2}$) Rydberg-molecular states on PECs for both Rb isotopes. The low value of $n$ leads to sub-$\%$ relative uncertainties and pronounced sensitivities to \mbox{$p$-wave} scattering. We have simultaneously fitted the \mbox{$s$-wave} and \mbox{$p$-wave} singlet and triplet scattering length functions $a_s^S(k)$, $a_s^T(k)$, $a_p^S(k)$, and $a_p^T(k)$. The binding energies depend on the functions $a_l^i(k)$ over a range of $k$, not only on $a_l^i(k \sim 0)$. The behavior of $a_p^T(k)$ near a shape resonance has a strong effect on states in the multiple-GHz-deep, inner wells around 600$a_0$. In future work, one may observe level splittings in these wells caused by the fine-structure splitting of the $^{3}P_{J=0,1,2}$ scattering channels~\cite{Khuskivadze2002praadiabatic,Eiles2017prahamiltonian}.

\section*{Acknowledgments}
We thank I. I. Fabrikant and Chris Greene for useful discussions. This work was supported by the NSF Grant No.~PHY-1506093. J.L.M. acknowledges support from the NSF Graduate Research Fellowship under Grant No. DGE 1256260.
%

\end{document}